# Characterization of Power-to-Phase Conversion in High-Speed P-I-N Photodiodes[1]

J. Taylor[1,2], S. Datta[3], A. Hati[1], C. Nelson[1], F. Quinlan[1], A. Joshi[3], and S. Diddams[1]

[1]*Time and Frequency Division, National Institute of Standards and Technology (NIST), Boulder, CO 80305, USA*

[2]*Department of Physics, University of Colorado, Boulder, CO, USA*

[3]*Discovery Semiconductors, 119 Silvia Street, Ewing, NJ 08628, USA*

*Corresponding authors: jennifer.taylor@nist.gov, scott.diddams@nist.gov*

Abstract: Fluctuations of the optical power incident on a photodiode can be converted into phase fluctuations of the resulting electronic signal due to nonlinear saturation in the semiconductor. This impacts overall timing stability (phase noise) of microwave signals generated from a photodetected optical pulse train. In this paper, we describe and utilize techniques to characterize this conversion of amplitude noise to phase noise for several high-speed (>10 GHz) InGaAs P-I-N photodiodes operated at 900 nm. We focus on the impact of this effect on the photonic generation of low phase noise 10 GHz microwave signals and show that a combination of low laser amplitude noise, appropriate photodiode design, and optimum average photocurrent is required to achieve phase noise at or below -100 dBc/Hz at 1 Hz offset a 10 GHz carrier. In some photodiodes we find specific photocurrents where the power-to-phase conversion factor is observed to go to zero.

**Index Terms**: frequency combs, microwave photonics, photodetectors

## 1. Introduction

Photodetection of pulsed or amplitude-modulated laser sources is one of the most basic components of photonic systems. Beyond ubiquitous lightwave communications systems, for some applications it is important to precisely know the arrival time of a light pulse, the timing structure of complex amplitude modulation on a pulse burst, or simply the rate of a periodic train of pulses. For example, photodetection of precisely timed signals is important for optical links used in emerging microwave photonics applications such as radio over fiber, phased-array radars [1], [2], arbitrary waveform generation [3], [4], radio astronomy [5], large-scale free-electron lasers [6], [7], and optical analog-to-digital conversion [8].

Our particular interest is the generation of ultra-low phase noise microwave tones and waveforms using a frequency-stabilized mode-locked laser comb [6], [7], [9]-[12]. For this application, we employ photodetection to convert the optical signals to electronic signals for use in and analysis with standard electronic devices. While additional noise contributions from components in the system can have a significant effect on the overall timing precision of these ultra-low noise signals, excess noise in the photodetection process cannot be neglected. Besides the fundamental shot noise of the photocurrent, the conversion of laser amplitude noise into electronic phase noise during photodetection has been previously identified as a limiting noise source for some applications [13]-[19]. We note that some alternative electro-optic techniques for extracting microwave signals from a mode-locked laser pulse train have recently been developed to address issues related to excess noise in photodetection [6], [20].

In this paper, we characterize the power-to-phase conversion that arises in part due to saturation and other nonlinearities in photodiodes driven by the pulse train from a mode-locked laser. Measurements are carried out on three InGaAs P-I-N photodiodes, described in Section 2, which are illuminated with the 1 GHz repetitive output of a mode-locked Ti:sapphire laser. We describe two techniques for measuring a 10 GHz amplitude-to-phase (AM-to-PM) coefficient, $\alpha$, that we define as the induced RMS phase variation (in radians) arising from a fractional power fluctuation ($\Delta P/P_o$). The techniques—which we term phase bridge and impulse response—independently measure power-induced phase fluctuations in the $10^{th}$ harmonic of the photodetected signal. These techniques are described in detail in Section 3, and the results obtained with the three photodiodes are presented. A comparison of these results as well as a discussion of the impact of AM-to-PM conversion on the RIN of a laser is given in Section 4.

The main outcomes of this work are: (1) The AM-to-PM coefficient $\alpha$ can be measured for specific photodiodes, and its dependence on operating parameters (bias voltage and average photocurrent) is quantified. Interestingly, $\alpha$ can be nulled for specific photocurrents in different photodiodes; (2) Design features in InGaAs P-I-N photodiodes that reduce the photocurrent density correspondingly reduce the AM-to-PM coefficient $\alpha$ over the broadest range of average photocurrents; (3) For a given photodiode detecting a frequency-stabilized repetitive pulse train, we can quantify the impact of laser relative intensity noise (RIN) on the phase noise of the resulting 10 GHz microwave signal. Knowing the laser RIN (measured in dBc/Hz) the predicted contribution of AM-to-PM to the single sideband phase noise is given by $L_{RIN}(f) = RIN + 20 \log(\alpha) - 3$ dB. Measured values of $\alpha$ at typical operating conditions fall in the range of 0.1 to 2 rad, implying, for example, that RIN must be lower than -77 dBc/Hz at 1 Hz offset to achieve L(f) below -100 dBc/Hz at the same offset; and (4) Operating at a null for $\alpha$ (e.g. $\alpha = 0$) opens the attractive possibility of eliminating the dependence of phase noise on laser RIN.

## 2. Experimental Setup

The three photodiodes have essentially identical internal semiconductor structure with different external features that affect saturation and photodiode performance (Figure 1a-c). The main structure is a p-side illuminated InGaAs photodiode with a 1 µm thick InP cap layer, which is transparent at wavelengths greater than 1000 nm; however, there is considerable absorption in this layer at 900 nm where the Ti:sapphire laser we employ operates. The photodiodes are packaged with coupling to a short (<1 m) step index fiber pigtail having ~9 µm core. Additionally, they are internally terminated with a 50 Ω matching resistor and reverse biased up to 9 V. The output microwave connection is via a coaxial (SMA) connector. The basic setup for connecting the photodiodes to the laser source is shown in Figure 1d. A concise description of the main differences of the three photodiodes is presented in Table 1.

The first photodiode, PD1, is a 22 GHz bandwidth, 30 µm photodiode with responsivity of 0.3 A/W at 900. The internal structure of the second photodiode, PD2, is identical to PD1; its unique feature is an integrated graded-index (GRIN) lens at the end of its SMF coupling fiber (see Figure 1b). The GRIN lens shapes the optical beam to produce a more uniform illumination profile (flat-top rather than Gaussian) on the photodiode; this more effectively illuminates the whole photodiode and suppresses peak photocurrent density [21], [22]. In the spectral range of 900 nm, the responsivity is 0.26 A/W. The third photodiode, PD3, is 60 µm in diameter and features a 0.3 µm InP cap layer. The combination of larger diameter and thinned cap layer improve the responsivity (measured to be 0.34 A/W) and power handling of this photodiode.

For the characterization of power-to-phase conversion in the photodiodes, the optical signal is generated from a 1 GHz repetition rate Ti:sapphire laser with a spectral peak centered around 900 nm containing 70 % of the total power [23], [24]. The bandwidth below 880 nm contains 25 % of the power (Figure 2). As already noted, the 900 nm wavelength is not ideally matched to maximize the responsivity and the bandwidth of the InGaAs photodiodes; however, this laser is employed in these tests because it is the source that has been previously used in the generation of low phase noise microwave signals [9], [10]. Moreover, we have found that in spite of the reduced responsivity relative to illumination at 1550 nm, the 10 GHz microwave power obtained after photodetection of this source exceeds that which is achieved with ~100 MHz repetition rate mode-locked Erbium fiber lasers.

The free space beam is coupled into a one meter SMF patch fiber, which is then coupled to the individual photodiode's fiber pigtail (Figure 1d). The pulse length measured with an autocorrelator at the end of this patch fiber is approximately 1 ps, and was verified to broaden only slightly after propagation in an additional meter of fiber. Optical power of the free space beam is controlled using a metal-on-glass variable neutral density filter wheel, causing minimal time and phase delay on the signal. Optical power is monitored at the end of this patch fiber with a power meter, and photocurrent can be determined using this power and the responsivity of the photodiode, or by monitoring the DC voltage at the output of the photodiode.

The photodiode converts the short optical pulse into an electrical pulse with duration given by the electrical bandwidth of the device. The illumination with the periodic optical pulse train gives rise to a comb of modes in the microwave domain separated by the repetition rate of the laser; in this case, the 1 GHz repetition rate produces a comb at 1 GHz, 2 GHz, 3 GHz, and so on up to the bandwidth of the photodiode. The desired microwave frequency, 10 GHz in our case, is electrically filtered and used as a frequency source. With illumination as just described, we obtain maximum 10 GHz signal powers near saturation of approximately -25 dBm at 1 mA for PD1, -13 dBm at 5 mA for PD2 and -6 dBm at 6 mA for PD3 [18]. Increasing the photocurrent above these values results in minimal increase of the achievable 10 GHz power, or even a decrease in some cases.

## 3. Measurement Methods

As mentioned in Section 1, we define an AM-to-PM coefficient, $\alpha$, as the induced RMS phase variation arising from a fractional power fluctuation ($\Delta P/P_o$). While $\alpha$ could be generally defined and measured for any frequency, we restrict most of our attention to the 10 GHz harmonic of the generated photocurrent. In order to calculate this coefficient, one must measure RMS phase fluctuations on the microwave signal from the photodiode in response to power modulation on the incident light. Here, we present two measurement techniques, which are described in detail below.

The primary technique we have studied employs a microwave phase bridge that permits the direct measurement of phase fluctuations relative to a stable 10 GHz reference with low uncertainty and high resolution (limited ultimately by fundamental thermal and shot noise). Variations of this well-developed technique have been previously applied to the measurement of saturation effects and AM-to-PM in photodiodes [13]-[15], [25] as well as intrinsic 1/f noise [14], [17]. The phase bridge approach has the advantage of being a high precision measurement, capable of measurement at the thermal noise limit with sub-femtosecond timing precision in narrow resolution bandwidths. However, this comes with the cost of additional complexity. We consider an alternative approach that employs more straightforward time-domain acquisition of data with an oscilloscope. In this case, we measure the time-domain electrical impulse response of the photodiode with a high-speed sampling oscilloscope and use Fourier analysis to extract power-dependent phase shifts at a desired frequency [16]. A time-domain measurement has the benefit of a simplified measurement system, but it comes with the possibility of increased noise and ambiguity in data analysis. This approach measures the power-dependent phase, but requires differentiation to obtain the power-to-phase coefficient $\alpha$. It further requires the use of a high-speed sampling oscilloscope, which is a well-developed laboratory instrument, but whose intrinsic timing jitter also introduces measurement uncertainties. We note that approaches have been developed to measure and remove time-base errors in sampling oscilloscopes, yielding precision at the level of ~200 fs [26-28]. The implementation of such time-base correction algorithms would likely decrease the uncertainty of the time-domain measurement presented here. Nonetheless, as will be shown, the results of the impulse response measurement agree with the phase-bridge measurement within a factor of two or less, and lend increased confidence to the phase bridge results. A more detailed comparison of the measurement results will be given below.

### a. Phase Bridge

We consider the phase bridge method the primary method for measuring power to phase conversion. A phase bridge is a microwave technique that allows sensitive relative measurement of phase and amplitude noise [29], [30]. Variations of this method relating to noise in photodiodes have been demonstrated previously [13]-[15], [17]. The phase bridge used here is most similar to that found in [15] with the main difference being the control of the laser repetition rate versus the reference synthesizer frequency. The block diagram of our system is shown in Figure 3. The repetition rate ($f_{rep}$) of the Ti:sapphire laser [23], [24], which is around 1 GHz, is mixed with a 1 GHz signal from a hydrogen maser-referenced synthesizer. This beat signal is fed into a loop filter servo and used to phase lock $f_{rep}$ to the synthesizer frequency using piezo-based control of the laser cavity length. The carrier-envelope offset frequency, $f_o$, of the Ti:sapphire laser can also be controlled [24], although it is typically left unlocked for this measurement. A portion of the Ti:sapphire laser output is sent through an acousto-optic modulator (AOM), which adds a known 20 kHz amplitude modulation on the laser light. Since AM-to-PM conversion in the photodiode changes some amount of this amplitude modulation into phase modulation, we monitor this spectral tone to quantify this conversion. The modulated beam goes through a neutral density (ND) filter wheel that is used to control the average optical power on the photodiode. At this point, the beam is coupled into fiber, which is then connected to the fiber pigtailed photodiode, as shown previously in Figure 1d. As we will discuss in Section 4, it is important to note that changing the fiber configuration in the system can affect AM-to-PM as well.

A bias tee at the output of the PD allows us to monitor the average DC voltage drop across the 50 Ω load resistor (and thus the photocurrent generated by the photodiode) on a high-impedance voltmeter while allowing frequency components of the signal >80 kHz to pass through to the rest of the measurement system. The 20 kHz tone from the AOM can also be monitored from the low frequency port of the bias tee. The high frequency signal is bandpass filtered around the tenth harmonic at 10 GHz. Following the filter, a combination of a microwave attenuator and low noise amplifier is used to keep the power of the signal at the LO port of the mixer constant while the optical power is being attenuated using the ND filter wheel. This ensures a constant mixer gain factor of $k_d$ = 0.47 V/rad, allowing us to convert voltage fluctuations to phase fluctuations.

The reference arm of the phase bridge is from a second maser-referenced synthesizer with its frequency set to match the 10 GHz signal in the test arm. A second photodiode using the same pulsed signal is not used in order to isolate only the AM-to-PM of one photodiode. Proper tuning of the relative phase between signals at the RF and LO ports of the mixer results in amplitude-insensitive detection. This is accomplished by adding a pure 30 kHz amplitude modulation to the amplified signal using a signal generator and a passive amplitude/phase modulator [31]. The phase shifter is then adjusted so that the 30 kHz tone, as seen on the FFT analyzer, is minimized. We note that this does not necessarily correspond to a minimum in voltage of the mixer output [32]. At this point, the measurement system is sensitive only to phase modulation in the test signal arising from any photodiode amplitude-to-phase conversion of the 20 kHz modulation on the light.

The RMS voltage of the initial 20 kHz modulation is measured on the FFT analyzer from the output of the bias tee and then divided by the average voltage to yield the normalized amplitude noise. The RMS voltage of the 20 kHz tone from the output of the mixer, which contains any phase modulation from AM-to-PM conversion in the PD, is then measured on the FFT analyzer and converted to radians using the gain factor $k_d$. The ratio of the two gives us the AM-to-PM conversion factor *α* for this method.

Figure 4 shows the results for all three photodiodes using the phase bridge method. The plots shown are averages of multiple measurements. PD2 and PD3 show consistently lower *α* across a wide range of photocurrent than PD1. This is consistent with the design parameters of these photodiodes (described in Fig. 1 and Table 1), which were implemented to reduce current density, saturation and related nonlinear effects. The impulse response measurements shown in the following section further confirm this higher power handling capability of PD2 and PD3 relative to PD1. The implication of these measurements is that reduction in photodiode nonlinearity indeed results in reduced AM-to-PM conversion. We note that low AM-to-PM at the largest photocurrent is especially desirable from the point of view of minimizing the shot-noise floor of generated microwave signals [18], [33]. In this regard, $\alpha$ for PD3 is below 0.4 rad up to a photocurrent of nearly 8 mA. The dashed green line indicates the measurement floor of the system; that is, how much AM-to-PM is present in the microwave components without the photodiodes. This is determined by using a 10 GHz synthesizer in place of the laser and photodiode in the phase bridge and measuring the AM-to-PM at varying signal power.

The most striking feature in the data of Figure 4 is the presence of nulls in the measured $\alpha$ of PD1 and PD2, indicating photocurrents where AM-to-PM conversion approaches zero. Such nulls were observed in earlier measurements [15], and at that time were attributed to impedance mismatch in the microwave components following the photodiode. However, we no longer believe this to be the case, as we consistently observe the nulls near the same average photocurrent irrespective of the exact lengths of microwave cables and the configuration of other components. It is important to note that our measurements do not determine the sign of the phase response to amplitude modulation (e.g., whether the 10 GHz phase is advanced or delayed with increasing power). Thus, the nulls are more correctly interpreted as zero crossings in the AM-to-PM response of the photodiode. From a physical picture, this implies that there are competing power-dependent processes related to the time delay of the removal of photocarriers, which apparently precisely cancel each other at specific photocurrents [13], [34], [35]. The advantage of operating the photodiode at one of these nulls is discussed in greater detail in Section 4.

### b. Impulse Response

Using Fourier analysis, we can extract phase information from the electrical impulse response of a photodiode to examine phase fluctuations that occur due to increasing the optical power incident on the photodiode [16]. Figure 5a shows the measurement setup for obtaining the electrical impulse response from the three photodiodes. These measurements employed a 20 GHz sampling oscilloscope, which is triggered by a second photodiode on which the optical power is kept constant. As already noted, timing errors intrinsic to sampling oscilloscopes have been studied and could be taken into account to improve this method [26-28]. Additionally, waveforms measured on the oscilloscope with and without a 10 dB attenuation between the photodiode and scope have the same shape, indicating that any nonlinearity in the amplitude of the scope response is small [36]. Since we are interested in comparing the results in general, more detailed analysis of these effects will not be presented here but would be interesting to address in future work.

The results are shown in Figure 5b-d. As the optical power is increased, a phase delay is evident by a time shift of the pulse's center of mass. Note that the effect is larger for PD1 (Fig. 5b) and PD2 (Fig. 5c) than for PD3 (Fig. 5d). As observed by others [15], [16], [25], this confirms that the structural differences in the photodiodes decrease the peak current density and reduce the saturation-induced delays, which give rise to AM-to-PM conversion.

The Fourier transform of the impulse response provides the RF phase in radians for a given Fourier frequency. Repeating this analysis for different average photocurrents yields the data of Figure 6a. The constant phase difference between the harmonics is largely arbitrary, arising from the impulse function being offset in time at the sampling oscilloscope (see Figure 6a); however, variations of the phase with photocurrent contain information about the AM-to-PM conversion. A cubic polynomial is fitted to the RF phase versus optical power, and the power-to-phase conversion

(PPC) factor, $\frac{d\varphi}{dP}$ [rad/W], at a given optical power is the derivative of the fitted line with respect to average optical power at that point. An absolute value is taken, though in principle the sign of the relative phase could be determined. The AM-to-PM coefficient, $\alpha$, is generated by normalizing the derivative at each point with the power $P_0$ at that point. Figure 6b demonstrates this analysis for three harmonics (1,3 and 10 GHz) for PD3 at a reverse bias $V_b$ = 9 V. While only three harmonics are shown here, this approach has the benefit of providing information about all harmonics generated by the photodiode up to its cutoff frequency.

The results of the impulse response method for all three photodiodes are shown in Figure 7. Figure 7a shows $\alpha$ calculated using the cubic polynomial fits, which approximate the shape of the phase changes and smooth possible noise in the data. A better fit and higher resolution can be obtained by taking more points close together (that is, changing optical power by smaller steps). The data go through one or two nulls and then rise to 1 rad or more at higher photocurrents. These nulls, given by derivatives of the cubic polynomial at an inflection point, were also seen in the direct measurement of α with the phase bridge method for PD1 and PD2 (see Figure 4).

In addition to using polynomial fits for derivatives, one also can generate the slope from a point-by-point derivative; that is, taking the "slope" of each pair of consecutive data points. This better represents the natural rise or fall of the phase with respect to power; however, this method is more susceptible to sudden, noisy jumps in the data. This method is shown in Figure 7b. We also see a large spike in $\alpha$ for PD1 in Figure 7b near 4 mA, which corresponds almost exactly to the peak and nulls seen in Figure 4 with the phase bridge method. However, the relative $\alpha$ levels for the three photodiodes are two times the levels given by the phase bridge method. This is likely due to the sensitivity of the point derivative method to jumps in the data.

To help minimize these jumps, one can also apply various types of numerical filters to smooth point derivatives. An in-depth analysis of these filters will not be presented in this paper, but we briefly show the results of a simple triangle filter in Figure 7c. We do see a reduction of the magnitude of $\alpha$ to more closely reflect the phase bridge results to a factor of 1.5, while maintaining much of the general structure of the point derivative method.

## 4. Results and Comparisons

In this section, we discuss the results and implications of the measurement of AM-to-PM conversion in the three photodiodes. The phase bridge and impulse response methods of measuring $\alpha$ show agreement within a factor of two or less, in some cases matching exactly. As already noted, while the phase bridge technique is considered to be more precise, the impulse response measurement has the advantage of simplified data acquisition. However, in our present implementation, this comes at the expense of increased noise and greater ambiguity in data analysis.

In principle, operating at a null for $\alpha$ would be ideal for experiments focused on the generation of low phase noise microwaves. This is especially the case for a null at a higher photocurrent. From the phase bridge measurement, there are two nulls for PD1 within the measured range of photocurrent; however, we have observed that these nulls can move to higher or lower photocurrent depending on specific operating parameters. For example, external factors such as temperature, bias voltage, and the shape of the optical mode of the light out of the fiber can impact the exact position of the nulls. We observed displacements of the nulls about half a milliamp by changing the external temperature by 10 °C or the bias voltage from 9 V to 5 V. At times, this shift can be large enough such that a photocurrent that once resulted in a null now results in $\alpha$ sitting at or near a peak.

Figure 8 specifically illustrates the effect of optical mode shaping. Because the fiber pigtail on the PD1 photodiode is multimode at our 900 nm operating wavelength, the light can occupy several higher order modes. By tightly coiling the fiber in loops of approximately 1 cm in diameter, these higher order modes can be reduced, thereby changing the illumination of the photodiode. While the mode structure was not rigorously analyzed, this simple example illustrates that changes in the spatial illumination of PD1 can move the placement of nulls by almost 0.5 mA, while the overall structure of $α$ versus photocurrent is unchanged. Of course, a potential variation such as this can be largely solved by employing fiber which is purely single mode at the illumination wavelength.

Another important result of quantifying AM-to-PM conversion is the ability to predict the impact of laser relative intensity noise (RIN) on the phase noise of a 10 GHz microwave signal from a frequency-stabilized repetitive pulse train. Knowing the laser RIN (measured in dBc/Hz), the predicted contribution of AM-to-PM to the single sideband phase noise is given by $L_{RIN}(f) = RIN + 20 \log(α) - 3$ dB.

To test the predictive utility of our measurements, we did a residual phase noise measurement on the 10 GHz harmonic produced with PD1 and PD2 and compared that to what we would predict based on the laser RIN and measured values of $α$. The RIN of the Ti:sapphire laser employed in this work is shown in Figure 9a. It was measured with the offset frequency unlocked, causing the high AM from the pump laser in the 100 Hz to 1 kHz region to be visible. From Figure 4, the "worst case" AM-to-PM values for PD1 and PD2 are 2.3 rad and 0.6 rad, respectively, at 4 mA. Using these values and the RIN of Figure 9a, we calculated the anticipated phase noise (Figure 9b) [18]. We then performed a residual phase noise measurement on the 10 GHz harmonic from each photodiode (see inset of Figure 9b). After the ND filter, a 50/50 fiber splitter sends half of the laser signal to the photodiode under test (PD1 or PD2) and half to a reference photodiode with a low, flat AM-to-PM coefficient. To insure the measurement is not limited by the noise floor, the laser power is adjusted so the photodiodes are at the "worst case" photocurrent. The rest of the two-arm set-up is similar to the phase bridge described in Section 3a. We compare those measurements to the predicted curves (Figure 9b). As seen, there is very good agreement between the estimate and the measurement in the high-AM region. Clearly, a good choice of photodetector and optimal operating parameters will ultimately reduce the phase noise. Using the above analysis, we can estimate the required laser RIN to achieve desired phase noise levels. For example, if the goal were to achieve 10 GHz microwave signals with single sideband phase noise less than -100 dBc/Hz at 1 Hz offset (corresponding approximately to fractional frequency instability of $1 \times 10^{-15}$), then the laser RIN would need to be at least as low at -83 dBc/Hz at the same offset if $α = 0.2$ rad. Operating any photodetector at a null has the potential to significantly reduce the impact of laser RIN on the microwave phase noise. In fact, at least one measurement employing an opto-electronic oscillator operating at 1300 nm has verified this assertion [19]. However, the dependence of the AM-to-PM coefficient on optical power (photocurrent) in that case appears to be different than our measurements. Further measurements to study the wavelength dependence of the AM-to-PM coefficient as well as investigations of the stability of the null to environmental conditions are ongoing and will be presented in a future publication.

## 5. Conclusions and Future Work

The conversion of amplitude fluctuations into phase fluctuations is a known effect in photodiodes but the conversion level varies among photodiodes and with photocurrent. For applications that require ultra-low noise microwaves generated from optical sources, the conversion from optical signals to electronic signals can introduce unwanted phase noise due to this effect. We have shown that the AM-to-PM coefficient $α$ for various photodiodes can be measured using different techniques, with results agreeing to within a factor of two or less. We show two design features of InGaAs P-I-N photodiodes that reduce peak current density and associated nonlinearities, thereby reducing the amplitude to phase coefficient $α$ over the broadest range of average photocurrents. Measurements of $α$, as shown here, together with appropriate design of the photodiode structure and its operating photocurrent can decrease the impact of power to phase conversion. One can also choose to operate at a photocurrent where there is a null or point of lower $α$ to minimize the conversion of amplitude noise from the source into excess phase noise [19]; however, nulls have been observed to move due to external factors and may be challenging to locate and precisely control. While the results presented here will be useful for experiments in the 900 nm spectral range, we anticipate the need for similar data and measurements with sources and photodiodes in the technologically important 1550 nm range.

## 6. Acknowledgements

The authors thank Paul Hale and William Swann for their helpful comments on this manuscript and Leo Hollberg for his efforts to initiate this work. Financial support was provided by NIST and a NIST SBIR grant to Discovery Semiconductor. This work is a contribution of an agency of the US government and not subject to copyright in the US.

# Figure Captions

**Figure 1**. Basic PIN structure and external modifications for a. PD1, b. PD2 with GRIN lens, and c. PD3, thinned InP cap layer. d. Coupling of free-space beam into photodiode through optical attenuator, fiber patch cable, and photodiode pigtail.

**Table 1.** Brief description of photodiodes and their notable structures.

**Figure 2**. Optical spectrum of the Ti:sapphire laser; the spectral peak centered around 900 nm containing 70 % of total power, and the bandwidth below 880 nm contains 25 % of the power.

**Figure 3**. Block diagram of phase bridge measurement set-up.

**Figure 4**. Average power-to-phase conversion factor, $α$, from phase bridge method. PD1 is at 7 V bias voltage, and PD2 and PD3 are at 9 V bias voltage.

**Figure 5**. a. Block diagram of impulse response measurement. Optical power is adjusted with the ND filter wheel, and impulse response is measured on the sampling oscilloscope. b-d. Impulse responses for the three photodiodes in increasing optical power steps (note optical power can be converted to photocurrent by multiplying by responsivity given in Table 1). PD1 at bias voltage of 7 V and PD1 and PD2 at bias voltage of 9 V. Note that as incident power increases, the pulses center shifts. This is most notable for b. PD1 and c. PD2. Not only does PD3 (d) minimize changes in pulse shift, it also generates a higher output voltage with a given optical power.

**Figure 6**. Fourier method as shown for PD3 at 9 V. (a) Phase (dots) is extracted from the Fourier transform of the impulse response at three Fourier frequencies, with cubic polynomial fits (solid lines). (b) Taking the derivative of phase with respect to power and then normalizing each point with its given optical power ($P_0$) gives $α$.

**Figure 7**. Compilation of magnitude of $α$ for the three photodiodes using a. impulse response with cubic fits, b. impulse response with point derivatives, c. impulse response with triangle filter.

**Figure 8**. Phase bridge results for PD1 with changes in shape of the optical mode incident on the detector due to the creation of ~1 cm loops in the optical fiber. Note large shifts of nulls and peaks.

**Figure 9**. a. AM spectrum of the free-running Ti:sapphire laser. b. Prediction of phase noise from laser AM spectrum (a). The "worst case" performance of PD1 and PD2 from Figure 4 is multiplied by the AM spectrum in (a). The residual phase noise is measured at a photocurrent where PD1 and PD2 exhibit high AM-to-PM conversion (around 4 mA in Figure 4). Inset: residual phase noise setup. Photodiode A is the photodiode under test; Photodiode B is the reference photodiode with low AM-to-PM.

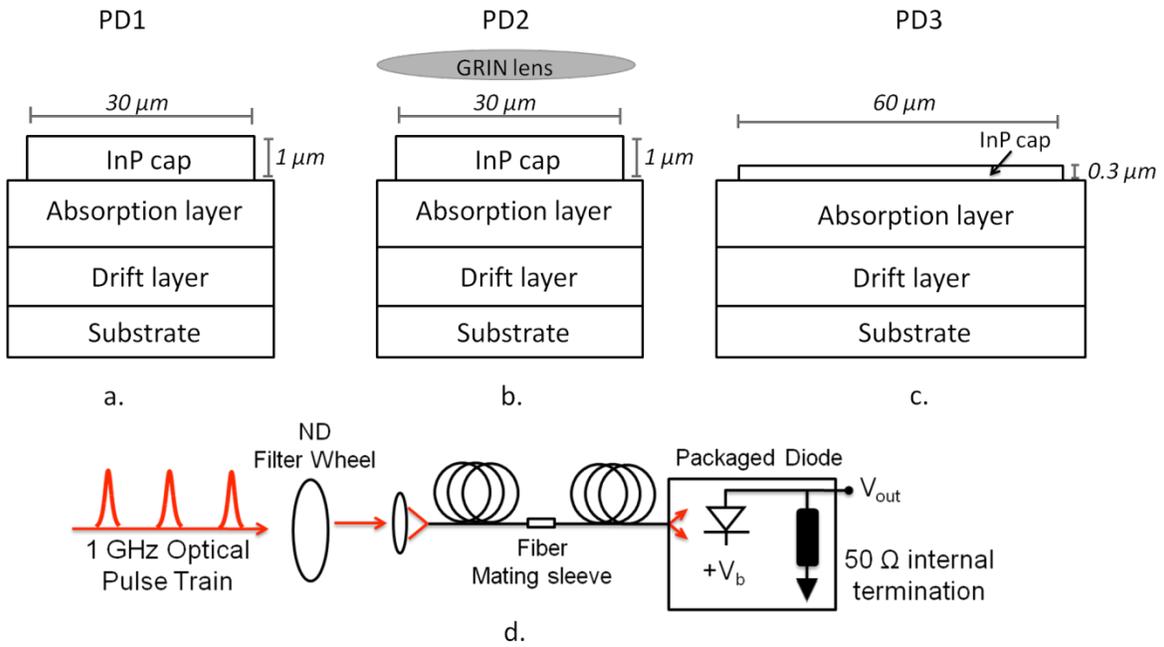

Figure 1

| PD | PD Diameter | Bandwidth | Responsivity @ 900 nm | Notable Structure |
|---|---|---|---|---|
| PD1 | 30 microns | 22 GHz | 0.30 A/W | SMF fiber |
| PD2 | 30 microns | 22 GHz | 0.26 A/W | GRIN lens |
| PD3 | 60 microns | 12 GHz | 0.34 A/W | Thinned InP cap layer |

Table 1

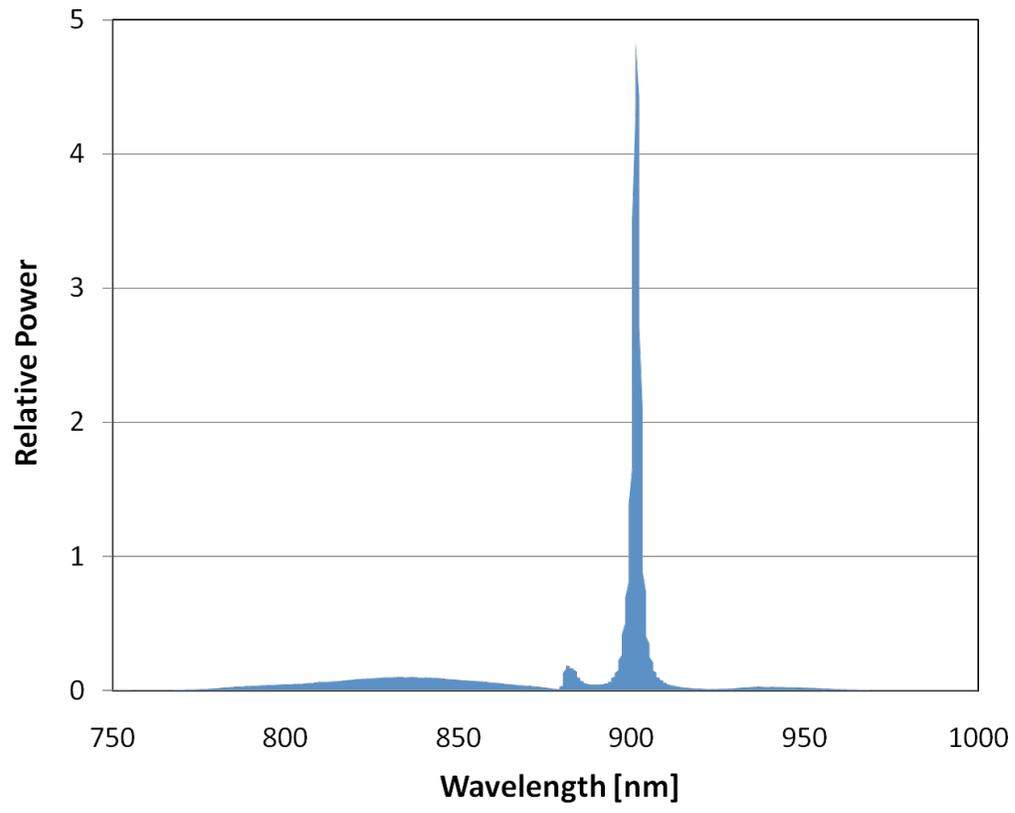

2
3
4
5
6
7
8
9
10   Figure 2
11
12

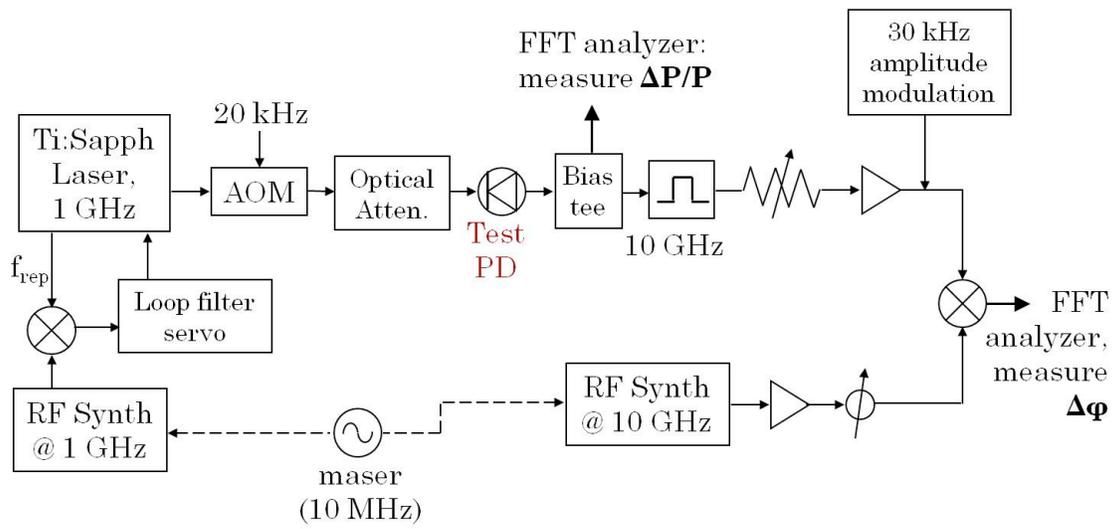

Figure 3

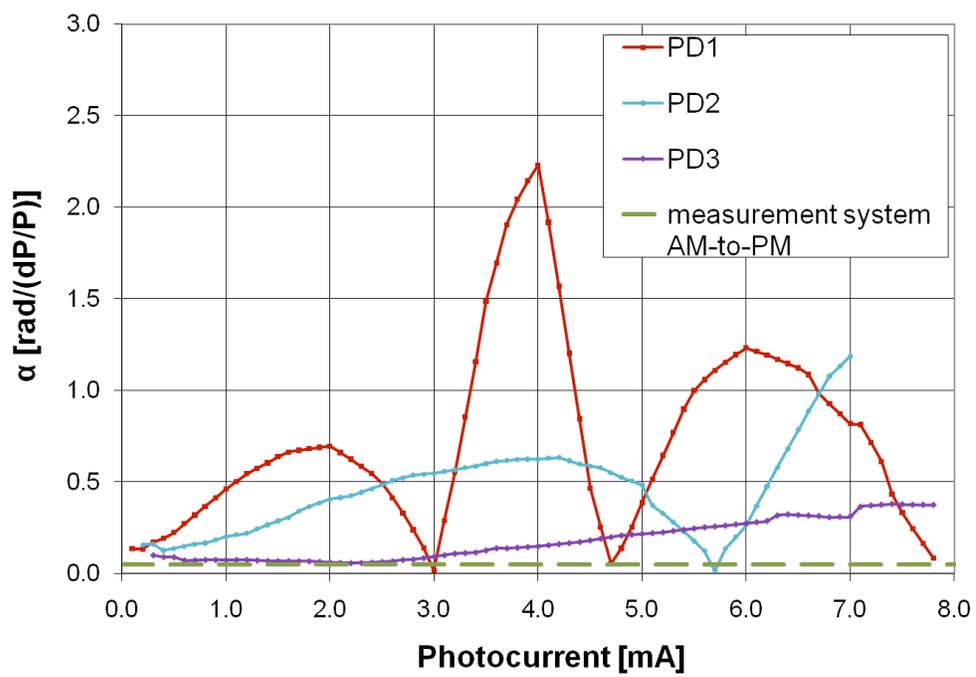

Figure 4

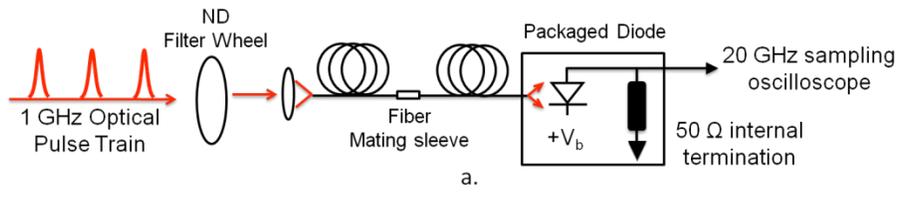
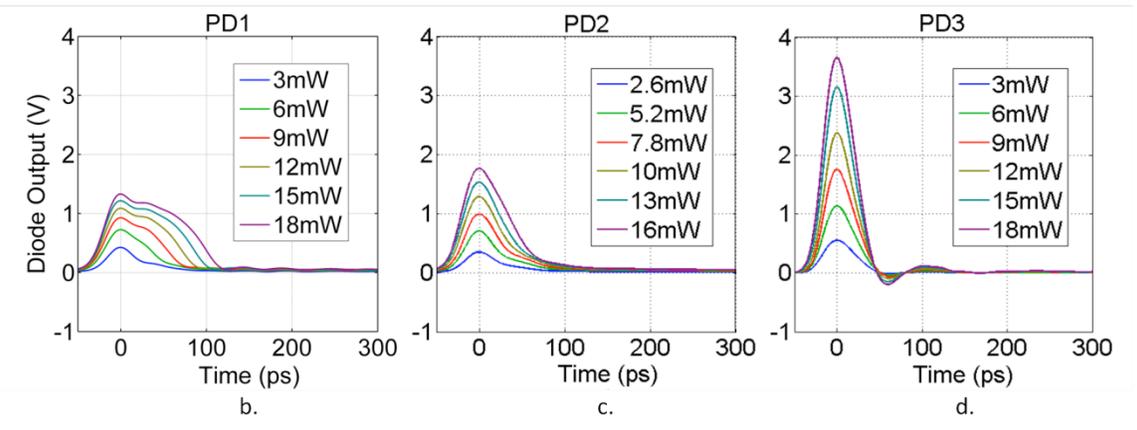

1 Figure 5

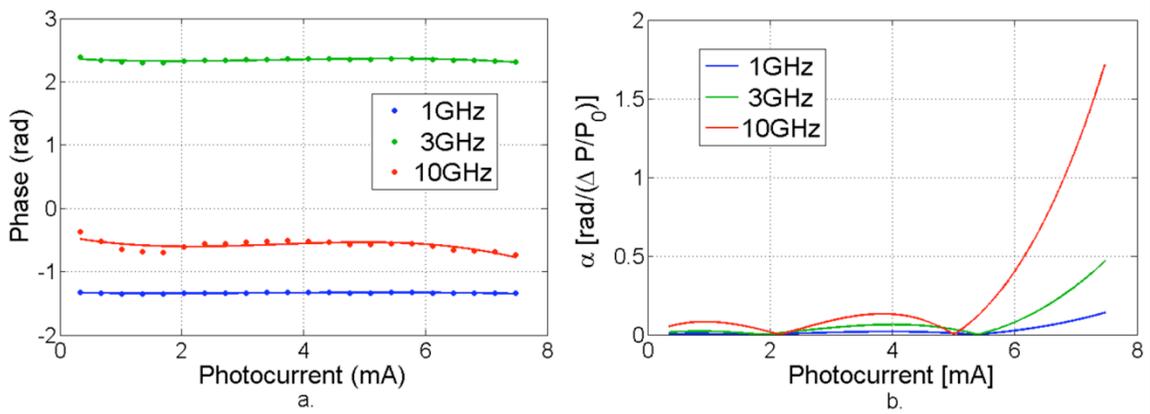

Figure 6

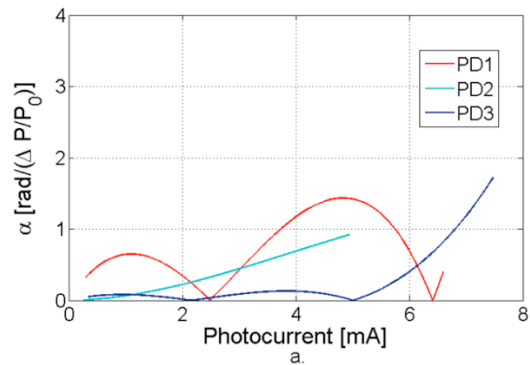
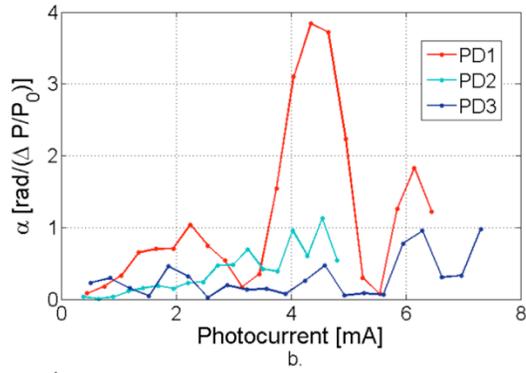
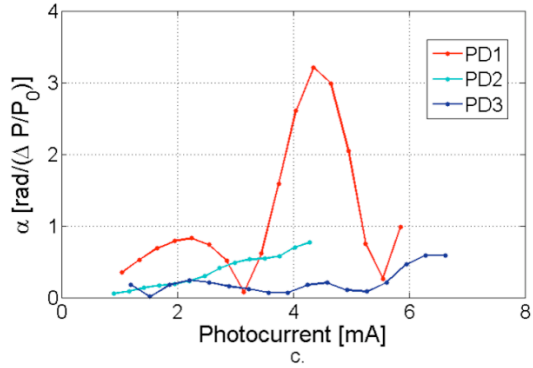

Figure 7

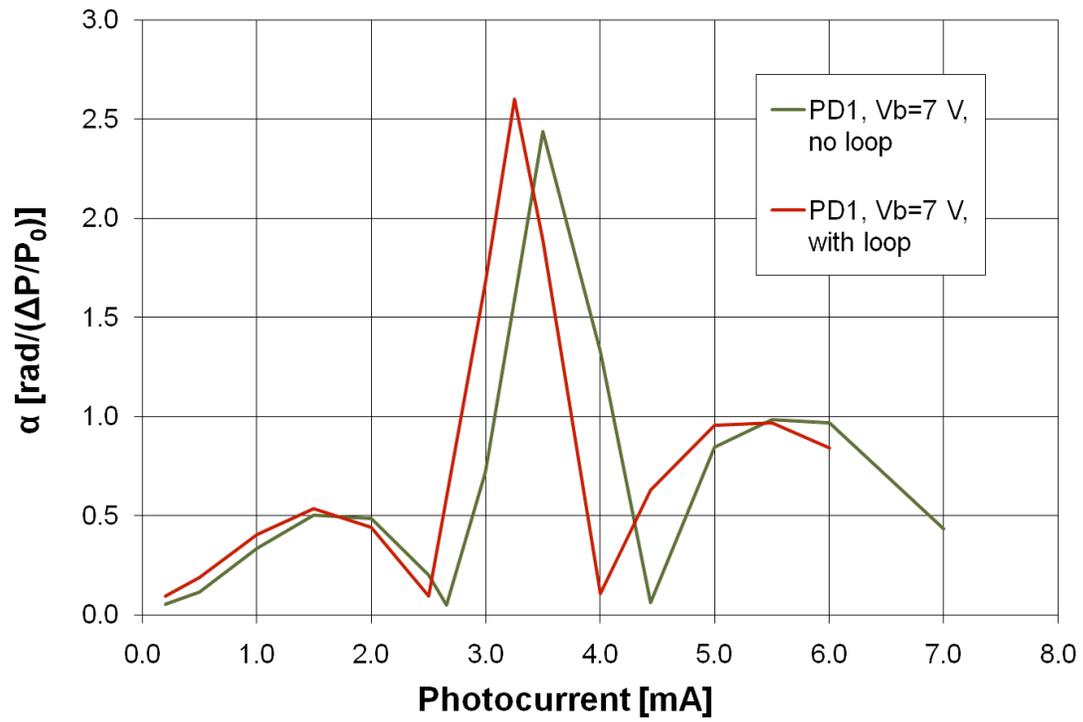

Figure 8

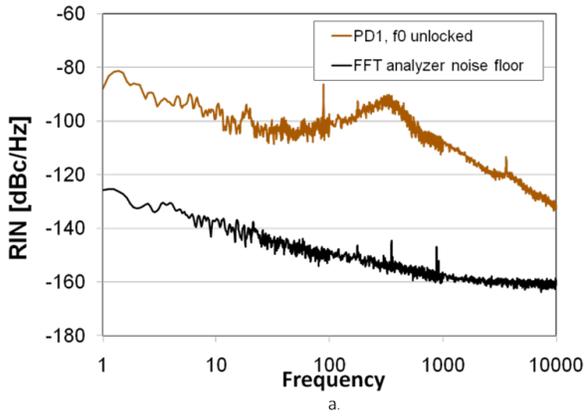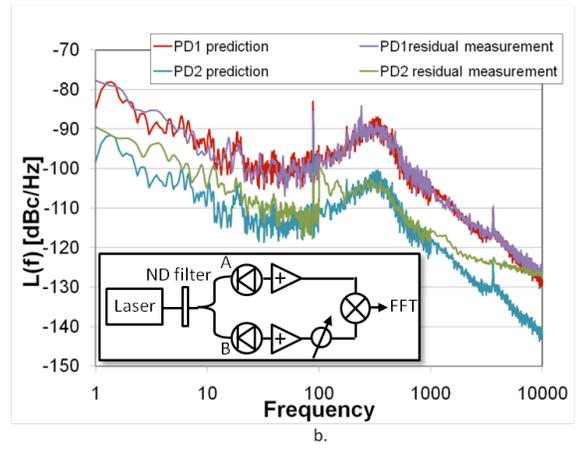

Figure 9